\documentclass[pra,amsmath,amssymb,reprint, nolongbibliography, superscriptaddress,nofootinbib,floatfix]
{revtex4-1} 
\usepackage{graphicx,helvet}
\usepackage{color}
\usepackage{bm}                      
\usepackage{float}
\usepackage{tikz}
\usepackage{braket}
\usepackage{soul}
\usepackage{amsfonts}
\usepackage{amsmath}
\usepackage{lipsum}
\definecolor{mygray}{gray}{0.4}
\definecolor{light-blue}{rgb}{0.8,0.85,1}
\graphicspath{{images/}}
\usepackage{amsmath}%
\usepackage{MnSymbol}%
\usepackage{wasysym}%
\usetikzlibrary{decorations.shapes}
\usepackage[draft,inline,nomargin]{fixme} \fxsetup{theme=color} 

\newcommand{\regens}{Institut f\"ur Theoretische Physik, Universit\"at Regensburg, D-93040 Regensburg, Germany}

\begin{document}

	\title{Interplay between coherent and incoherent decay processes in chaotic systems: \\ the role of quantum interference}
	\author{Camilo Moreno} \affiliation{\regens}
	\author{Juan-Diego Urbina} \affiliation{\regens}
	\email{camilo-alfonso.moreno-jaimes@stud.uni-regensburg.de}

\begin{abstract} \small
The population decay due to a small opening in an otherwise closed cavity supporting chaotic classical dynamics displays a quantum correction on top of the classical exponential form, a pure manifestation of quantum coherence that acquires a universal form and can be explained by path interference. Being coherent, such enhancement is prone to decoherence effects due to the coupling of the system to an external environment. We study this interplay between incoherent and coherent quantum corrections to decay by evaluating, within a Caldeira-Legget scenario, off-diagonal contributions to the decoherence functional coming from pairs of correlated classical paths in the time regime where dissipative effects are neglected and decoherence does not affect the classical dynamics, but quantum interference must be accounted for. We find that the competing effects of interference and decoherence lead to a universal non-monotonous form for the survival probability depending only on the universality class, coupling strength, and macroscopic parameters of the cavity.
\end{abstract}
\maketitle
\section{Introduction}
The very discovery of quantum phenomena with the progressive and unstoppable extension of the corresponding quantum domain into larger and larger scales is due to the ability to isolate physical systems from its environment \cite{Zurek2003}. Pretty much as Galileo was able to understand that despite daily intuition the natural state of motion is constant velocity instead of rest and changes are due to external influences \cite{Goldstein}, the founding fathers of quantum mechanics recognized that at the fundamental level an isolated system remains coherent thus displaying a behavior that is classically counter-intuitive.

This conceptual realization is the more impressive when one considers that, actually, the very act of observation unavoidably requires the pristine evolution of closed and isolated quantum systems to account for the interface between the system and the observation device \cite{Schlosshauer,Neumann}. In fact, the way we probe the most fundamental quantum aspects of closed systems, like the discreetness of the energy spectrum, is through scattering experiments where the system is coupled with a continuum: we coherently couple the system to its electromagnetic environment, and then couple the latter to a measurement device. As clarified by several decades of efforts, it is at this last stage where both the possibility of extracting information from the system and the corresponding lost of coherence, decoherence, takes place \cite{Zurek1981,Zurek1982, Experiment1,Experiment2,Expetiment3,Experiment4}.

The interplay between coherent decaying due to the opening of the system to a coherent continuum, as in scattering systems, and decoherence, usually modelled by coupling to a large set of uncontrolled degrees of freedom, takes a further twist if one is interested in studying such interplay in a regime of large systems or high quantum numbers, the so-called mesoscopic regime \cite{Klausbook}. In this case, the microscopic description takes advantage of the universal quantum signatures of systems with chaotic classical limit that are explored by means of asymptotic analysis based on path integrals \cite{Libro_chaos}. In this way, the interplay between quantum coherence, decoherence and quantum signatures of chaos is a pillar of modern physics, with broad applications, from the theory of quantum transport \cite{Jacquod-transport}, to the precise understanding of the quantum-classical transition \cite{Zurek1981}. 

In previous works the universal quantum corrections to classical decay in open chaotic systems were computed \cite{Daniel} in the spirit of the semiclassical approach to mesoscopic transport. Our objective here is to extend these ideas in a way that addresses the key impact of decoherence on such effects. In order to account for the emergence of universal quantum signatures of classically chaotic dynamics the proper tools are those of semiclassical analysis where quantum phenomena are described in terms of a highly non-trivial use of classical information around classical solutions. Specifically, quantum interference is explained in terms of interfering classical paths, and, as we will show here, its degrading due to decoherence is explained in terms of decoherence functionals evaluated themselves along pairs of classical solutions. A key finding of our analysis is that in the limit of weak coupling the leading classical contribution to the decoherence processes can be shown to vanish, and therefore all its effects arise from quantum interference, fully captured by the semiclassical theory of correlated solutions to produce universal results.

The paper is organized as follows. First, section \ref{Caldeira} presents the general aspects of decoherence due to the coupling of a particle to a bath reservoir within the Caldeira-Leggett model. Section \ref{survival} is devoted to review the main features of the quantum survival probability, which involves a scenario where the particle is inside a cavity from which it may escape. The main technical aspects of our work, where we develop the semiclasscial treatment of the particle inside a cavity coupled to a bath, and study the first quantum corrections in the semiclassical limit is the subject of section \ref{semicl}. Finally, we provide some concluding remarks in section \ref{conlc}.

\section{Decoherence in the Caldeira-Leggett model}\label{Caldeira}
Following the standard Feynman-Vernon approach as made explicit by the Caldeira-Legget model, we consider a particle $\mathcal{A}$ coupled to a $N$-particle environment of harmonic oscillators $\mathcal{E}$. The total Hamiltonian reads
\begin{equation}
\hat{H}=\hat{H}_{\mathcal{A}}+\hat{H}_{\mathcal{E}} +\hat{H}_{\mathcal{A}\mathcal{E}},
\end{equation}
where 
\begin{equation}
 \hat{H}_{\mathcal{A}}=\frac{\hat{\mathbf{P}}^2}{2m}+\hat{V}(\hat{\mathbf{Q}})   
\end{equation}
 is the Hamiltonian of the central system with $f>1$ degrees of freedom,
\begin{equation} 
 \hat{H}_{\mathcal{E}}=\sum_{k=1}^{N}\frac{1}{2}\Big(\hat{p}_k^2/m_k+m_k\omega_k^2\hat{q}_k^2\Big),
 \end{equation}
 the Hamiltonian of the environment, and $\hat{H}_{\mathcal{A}\mathcal{E}}$ the interaction energy between $\mathcal{A}$ and $\mathcal{E}$. 
We will choose an interaction which couples linearly the position operator of the central system $\hat{\mathbf{Q}}$ with the position operator of each particle of the environment $\hat{q}_k$, with $k=1,...,N$; which reads
\begin{equation}\label{interaction}
\hat{H}_{\mathcal{A}\mathcal{E}}=-\hat{\mathbf{Q}}\otimes\sum_{k=1}^{N}g_k\hat{q}_k+\hat{\mathbf{Q}}^2\sum_{k=1}^{N}\frac{g_k^2}{2m_k\omega_k^2}.
\end{equation}
The last term in Eq.~(\ref{interaction}) compensates for the coupled-induced renormalization of the potential \cite{GRABERT}.

While Eq.~(\ref{interaction}) will in general produce dissipation as well as decoherence on $\mathcal{A}$ \cite{Weiss}, in this paper we will consider the regime of "pure decoherence", neglecting  dissipative effects, an approximation that is fully justified due to the vast separation of time scales between these two mechanisms. That is, we will be only interested in the decoherence effects that $\mathcal{E}$ produces on the central system. 

The whole system $\mathcal{A}+\mathcal{E}$ evolves under $\hat{H}$, with the time evolution being described by the associated propagator given by
\begin{gather}
K(\mathbf{Q}_f,\mathbf{q}_f,t;\mathbf{Q}_i,\mathbf{q}_i)= \bra{\mathbf{Q}_f,\mathbf{q}_f}\mathrm{e}^{-\frac{i}{\hbar} \hat{H}t}\ket{\mathbf{Q}_i,\mathbf{q}_i},
\end{gather} 
with the vector $\mathbf{q}$ defined as $\mathbf{q}=(q_1,...,q_N)$.

In the Feynman path integral approach the propagator has the form
\begin{gather}
K(\mathbf{Q}_f,\mathbf{q}_f,t;\mathbf{Q}_i,\mathbf{q}_i)= \int \mathcal{D}[\mathbf{Q}(s),\mathbf{q}(s)]\mathrm{e}^{\frac{i}{\hbar} R[\mathbf{Q},\mathbf{q}]},
\end{gather}
which is a sum over all paths with boundary conditions: $\mathbf{Q}_i=\mathbf{Q}(0), \mathbf{Q}_f=\mathbf{Q}(t);\mathbf{q}_i=\mathbf{q}(0), \mathbf{q}_f=\mathbf{q}(t)$, and $R$ is the total action $R=R_{\mathcal{A}}+R_{\mathcal{E}}+R_{\mathcal{A}\mathcal{E}}$.

A general initial state $\rho_{\mathcal{A}\mathcal{E}}$ will evolve as
\small
\begin{equation*}
\begin{gathered}
 \rho_{\mathcal{A}\mathcal{E}}(\mathbf{Q}_f,\mathbf{Q}'_f,\mathbf{q}_f,\mathbf{q}'_f,t)= \\ \int d\mathbf{Q}_i  d\mathbf{Q}'_i d\mathbf{q}_i  d\mathbf{q}'_i K(\mathbf{Q}_f,\mathbf{q}_f,t;\mathbf{Q}_i,\mathbf{q}_i) K^*(\mathbf{Q}'_f,\mathbf{q}'_f,t;\mathbf{Q}'_i,\mathbf{q}'_i) \rho_{\mathcal{A}\mathcal{E}}, 
 \end{gathered}	
\end{equation*} 	
\normalsize
and the reduced dynamics of the central system is obtained after tracing out the degrees of freedom of the environment, $\rho_{\mathcal{A}}=\rm{Tr}_{\mathcal{E}}\Big[\rho_{\mathcal{A}\mathcal{E}}\Big]$. 

Choosing a factorized initial state $\rho_{\mathcal{A}\mathcal{E}}(0)=\rho_{\mathcal{A}}(0)\otimes\rho_{\mathcal{E}}(0)$, the reduced density matrix gives
\begin{equation}
\begin{gathered}
\rho_{\mathcal{A}}(\mathbf{Q}_f,\mathbf{Q}'_f,t)=\int d\mathbf{Q}_i d\mathbf{Q}'_i \rho_{\mathcal{A}}(0) \\ 
\int \mathcal{D}[\mathbf{Q}(s)]\mathcal{D}[\mathbf{Q}'(s)]
\mathrm{e}^{\frac{i}{\hbar}(R_{\mathcal{A}}[\mathbf{Q}]-R_{\mathcal{A}}[\mathbf{Q}'])}\mathcal{F}[\mathbf{Q},\mathbf{Q}'],
\end{gathered}
\end{equation} 
Where $\mathcal{F}[\mathbf{Q},\mathbf{Q}']$ is the Feynman-Vernon influence functional
given by
\begin{equation}
\begin{gathered}
\mathcal{F}[\mathbf{Q},\mathbf{Q}']=\int_{(\mathbf{q}_i, \mathbf{q}'_i) \to \mathbf{q}_f} d\mathbf{q}_f d\mathbf{q}_i d\mathbf{q}'_i \rho_{\mathcal{E}}(0) \\
\int \mathcal{D}[\mathbf{q}]\mathcal{D}[\mathbf{q}']\mathrm{e}^{\frac{i}{\hbar}\big(R_{\mathcal{E}}[\mathbf{q}]+R_{\mathcal{A}\mathcal{E}}[\mathbf{Q},\mathbf{q}]-R_{\mathcal{E}}[\mathbf{q}']-R_{\mathcal{A}\mathcal{E}}[\mathbf{Q}',\mathbf{q}']\big)}.
\end{gathered}
\end{equation}

If we choose an initial thermal state for $\mathcal{E}$ at inverse temperature $\beta=1/\kappa_B T$, $\hat{\rho}_{\mathcal{E}}=\frac{\mathrm{e}^{-\beta \hat{H}_{\mathcal{E}}}}{Z_{\mathcal{E}}}$, the influence functional has an exact solution \cite{GRABERT} given by
\begin{equation}\label{rho_A}
\begin{gathered}
\rho_{\mathcal{A}}(\mathbf{Q}_f,\mathbf{Q}'_f,t)=\int d\mathbf{Q}_i d\mathbf{Q}'_i \rho_{\mathcal{A}}(0) \\
\int \mathcal{D}[\mathbf{Q}]\mathcal{D}[\mathbf{Q}']\mathrm{e}^{\frac{i}{\hbar}\big(R_{\mathcal{A}}[\mathbf{Q}]-R_{\mathcal{A}}[\mathbf{Q}']-R^{F}[\mathbf{Q},\mathbf{Q}']\big)}
\mathrm{e}^{-R^d[\mathbf{Q},\mathbf{Q}']/\hbar}.
\end{gathered}
\end{equation} 

The effective action in $R^{F}$ is responsible for dissipation of energy of the particle and thus for the relaxation process. Neglecting this term in our approximation we only keep the decoherence action $R^d$, which reads
\begin{equation}
\begin{gathered}
R^d[\mathbf{Q},\mathbf{Q}']= \int_{0}^{t} ds\int_{0}^{s} du  \\
(\mathbf{Q}(s)-\mathbf{Q}'(s))^T\kappa(s-u)(\mathbf{Q}(u)-\mathbf{Q}'(u)),
\end{gathered}
\end{equation}
which involves a mixture of off-diagonal components $\mathbf{Q}-\mathbf{Q}'$ of the density matrix along paths mediated by the bath kernel 
\begin{equation}
\kappa(s-u)=\frac{1}{\pi} \int_{0}^{\infty}d\omega J(\omega)\coth\big(\beta \hbar \omega/2\big)\cos\omega(s-u). 
\end{equation}

Here the integral is over a continuum of bath-oscillators frequencies $\omega$. 
\newline 

If we choose an Ohmic spectral density associated with the bath such that $J(\omega)=\Gamma \omega$, in the limit of high temperatures, $\beta \to 0$, the kernel transforms into 
\begin{equation}
\kappa(s-u)=\frac{2\Gamma}{\hbar \beta}\delta(s-u), 
\end{equation}
and the decoherence term in the action takes the form   
\begin{equation}\label{decoh_action}
R^d[\mathbf{Q},\mathbf{Q}']=\frac{2\Gamma}{\hbar \beta} \int_{0}^{t} ds |\mathbf{Q}(s)-\mathbf{Q}'(s)|^2.
\end{equation}
In this way, inserting Eq.~(\ref{decoh_action}) into Eq.~(\ref{rho_A}), we see that $R^d$ is responsible for the suppression  of quantum coherence between paths $\mathbf{Q}$ and $\mathbf{Q}'$ due to the coupling of the central system to the environment of bath oscillators.

All together, in the purely coherent, high-temperature regime, Eq.~(\ref{rho_A}) reads
\begin{equation}\label{rho}
\begin{gathered}
\rho_{\mathcal{A}}(\mathbf{Q}_f,\mathbf{Q}'_f,t)=\int d\mathbf{Q}_i d\mathbf{Q}'_i \rho_{\mathcal{A}}(0) \\
\int \mathcal{D}[\mathbf{Q}]\mathcal{D}[\mathbf{Q}']\mathrm{e}^{\frac{i}{\hbar}(R_{\mathcal{A}}[\mathbf{Q}]-R_{\mathcal{A}}[\mathbf{Q}'])}
\mathrm{e}^{-\alpha \int_{0}^{t} ds |\mathbf{Q}(s)-\mathbf{Q}'(s)|^2},
\end{gathered}
\end{equation} 
where we have defined a new coupling-strength constant $\alpha$ subduing the whole parameter dependence of the decoherence action.

To make further progress, as Eq.~(\ref{rho}) represents still a formidable problem, we will assume that $\alpha$ is classically small so that the coupling with the environment does not affect the classical dynamics of the central system, only the coherence between pair of paths $\mathbf{Q}, \mathbf{Q}'$. This weak-coupling regime, usually justified even for realistic models, will enable us to evaluate Eq.~(\ref{rho}) semiclassically. 

Eq.~(\ref{rho}) is thus a model for decoherence without dissipation process. The main assumption is that the coupling with the bath is classically small such that the central system only experiences a loss of coherence of the relative states $\mathbf{Q}, \mathbf{Q}'$. This is also justified if we note that in these models the decoherence time scale is much more faster than the dissipative time scales induced by the environment \cite{Weiss,Zurekinbook}.

\section{Particle in a chaotic cavity and quantum survival probability}\label{survival}
In \cite{Gutierrezprl} the authors considered a particle moving in two dimensions, initially inside a cavity of area $A$. The cavity has a hole of size $l$ from which the particle can escape. 

At the purely classical level, it is known that the  probability $\rho_{\text{cl}}$ to find the particle inside the cavity at time $t$, the so-called survival probability, has the form \cite{Daniel}
\begin{equation}
\rho_{\text{cl}}=\mathrm{e}^{-t/\tau_D}, 
\end{equation}
for cavities supporting classical chaotic dynamics. This result is valid for times longer than the Lyapunov time $1/\lambda$, with $\lambda$ the Lyapunov exponent (assumed uniform). Here, $1/\tau_D$ is the escape rate, given in terms of the dwell time $\tau_D=\Omega(E)/(2lp)$, where $p$ is the momentum of the particle, and we introduced the phase-space volume of the energy shell $E$, $\Omega(E)=\int d\mathbf{Q}d\mathbf{P}\delta(E-H_{\mathcal{A}}(\mathbf{Q},\mathbf{P}))$.  

In \cite{Gutierrezprl}, using semiclassical techniques, quantum corrections to the classical survival probability were studied, and a universal quantum enhancement for underlying classical chaotic dynamics was predicted. At first order in $\hbar$, it takes the form of a correction $\delta \rho_{\text{qm}}$,  
\begin{equation}\label{qm_survival}
\delta \rho_{\text{qm}}= \mathrm{e}^{-t/\tau_D}\frac{t^2}{2T_H\tau_D},   
\end{equation}
where $T_H=\Omega/(2\pi \hbar)$ is the Heisenberg time.

This quantum enhancement of the decaying classical survival probability is a \textit{coherent} effect coming from interference between pair of trajectories as shown in \cite{Gutierrezprl}.

In the following, we will study the interplay between this quantum survival probability and the \textit{decoherence} process as implied by Eq.~(\ref{rho}). That is, we will consider a particle that is coupled {\it both} to a continuum through an opening of size $l$ of the cavity that produces coherent effects, and to an environment that suppresses such effects by decoherence.

\section{Semiclassical treatment}  \label{semicl} 

As reported for the first time in \cite{Gutierrezprl,Kuipers_2009}, in general coherent corrections to the classical dynamics of observables, like the survival probability, manifest themselves only when the observable itself is only defined within a finite region of an otherwise unbounded system. In this spirit, the state of the particle $\mathcal{A}$ inside the cavity under the influence of $\mathcal{E}$ will evolve using 
Eq.~(\ref{rho}), but projected onto the area of the open cavity. 

We implement the semiclassical approach to Eq.~(\ref{rho}) taking into account that, in our  weak-coupling scenario, the classical solutions of the saddle-point analysis (SPA) in Eq.~(\ref{rho}) are given by solutions of the classical equations of motion coming from the stationary condition of the bare action $R_{\mathcal{A}}$. This leads us to consider the application of SPA at the level of the amplitudes, the so-called semiclassical approximation to the quantum mechanical propagator \cite{Gutzwiller}. 

Within the semiclassical approximation, the propagator takes the form
\begin{equation}\label{propagator-van-vleck}
K_{\text{sc}}(\mathbf{Q}_f,t;\mathbf{Q}_i,0)= \frac{1}{2\pi\hbar}\sum_{\tilde{\gamma}: \mathbf{Q}_i \to \mathbf{Q}_f}A_{\tilde{\gamma}}\mathrm{e}^{\frac{i}{\hbar}R^{\tilde{\gamma}}_{\mathcal{A}}},
\end{equation}
as a sum over classical paths $\tilde{\gamma}$ connecting points $\mathbf{Q}_i \to \mathbf{Q}_f$, during time $t$. 
The van Vleck-Gutzwiller amplitude
\begin{equation}
A_{\tilde{\gamma}}=\Bigg|\rm{det}\Big(-\frac{\partial^2 R^{\tilde{\gamma}}_{\mathcal{A}}}{\partial \mathbf{Q}_f \partial \mathbf{Q}_i}\Big)\Bigg|^{1/2}\mathrm{e}^{-i\pi\mu_{\tilde{\gamma}}/2} ,
\end{equation}
contains, besides the stability factor, the number of focal points $\mu_{\tilde{\gamma}}$ of the trajectory.

Substitution of Eq.~(\ref{propagator-van-vleck}) into the general expression for evolution of the state in Eq.~(\ref{rho}), gives 
\begin{equation}\label{van_vleck}
\begin{gathered}
\rho^{\text{sc}}_{\mathcal{A}}(\mathbf{Q}_f,\mathbf{Q}'_f,t)= \frac{1}{(2\pi\hbar)^2}\int_A d\mathbf{Q}_i d\mathbf{Q}'_i \rho_{\mathcal{A}}(\mathbf{Q}_i,\mathbf{Q}'_i) \\
\sum_{\tilde{\gamma}: \mathbf{Q}_i \to \mathbf{Q}_f} \sum_{\tilde{\gamma}': \mathbf{Q}'_i \to \mathbf{Q}'_f} A_{\tilde{\gamma}}A^*_{\tilde{\gamma}'} \mathrm{e}^{\frac{i}{\hbar}(R^{\tilde{\gamma}}_{\mathcal{A}}-R^{\tilde{\gamma}'}_{\mathcal{A}})} 
\mathrm{e}^{-\alpha \int_{0}^{t} ds |\mathbf{Q}_{\tilde{\gamma}}(s)-\mathbf{Q}_{\tilde{\gamma}'}(s)|^2},
\end{gathered}
\end{equation}         
thus taking the form of a double sum over classical paths.

Since the semiclassical approximation in Eq.~(\ref{van_vleck}) is valid when the bare action of the central system $R_{\mathcal{A}}$ is much greater than $\hbar$,  the sum over pair of trajectories contains highly oscillatory terms that cancels  out each other, unless the pair difference is of order $\hbar$,  $R^{\tilde{\gamma}}_{\mathcal{A}}-R^{\tilde{\gamma}'}_{\mathcal{A}} \sim \mathcal{O}(\hbar)$.  Following the usual semiclassical methods \cite{Sieber_1999}, the important contributions to the double sum come from those pairs of trajectories close to each other in phase space. 

In the double sum of Eq.~(\ref{van_vleck}) we look for pair of trajectories $(\gamma,\gamma')$ which start at the same point $\mathbf{r}_i=(\mathbf{Q}_i+\mathbf{Q}'_i)/2$ and end at the same final point $\mathbf{r}_f=(\mathbf{Q}_f+\mathbf{Q}'_f)/2$, where now we define the relative and center of mass coordinates $\mathbf{y}=\mathbf{Q}-\mathbf{Q}'$, $\mathbf{r}=(\mathbf{Q}+\mathbf{Q}')/2$.

In a final step we expand the action $R^{\tilde{\gamma}}_{\mathcal{A}}$ around the path $\gamma$, 
\begin{equation}
R^{\tilde{\gamma}}_{\mathcal{A}}(\mathbf{Q}_i,\mathbf{Q}_f) \approx  
R^{\gamma}_{\mathcal{A}}(\mathbf{r}_i,\mathbf{r}_f)-\mathbf{P}^{i}_{\gamma}\cdot \mathbf{y}_i/2+\mathbf{P}^{f}_{\gamma}\cdot \mathbf{y}_f/2,
\end{equation}
and similarly $R^{\tilde{\gamma}'}_{\mathcal{A}}$ around $\gamma'$, thus Eq.~(\ref{van_vleck}) reads now 
\small
\begin{equation}\label{van_vleck_1}
\begin{gathered}
\rho^{\text{sc}}_{\mathcal{A}}(\mathbf{r}_f+\mathbf{y}_f/2,\mathbf{r}_f-\mathbf{y}_f/2,t)=\int_A d\mathbf{r}_i d\mathbf{y}_i \rho_{\mathcal{A}}(\mathbf{r}_i+\mathbf{y}_i/2,\mathbf{r}_i-\mathbf{y}_i/2) \\
\frac{1}{(2\pi\hbar)^2}\sum_{\gamma: \mathbf{r}_i \to \mathbf{r}_f} \sum_{\gamma': \mathbf{r}_i \to \mathbf{r}_f} A_{\gamma}A^*_{\gamma'} \mathrm{e}^{\frac{i}{\hbar}(R^{\gamma}_{\mathcal{A}}-R^{\gamma'}_{\mathcal{A}})} \mathrm{e}^{-\frac{i}{\hbar}(\mathbf{P}^{i}_{\gamma}+\mathbf{P}^{i}_{\gamma'})\cdot \mathbf{y}_i/2} \\
\mathrm{e}^{\frac{i}{\hbar}(\mathbf{P}^{f}_{\gamma}+\mathbf{P}^{f}_{\gamma'})\cdot \mathbf{y}_f/2} 
\mathrm{e}^{-\alpha \int_{0}^{t} ds |\mathbf{r}_{\gamma}(s)-\mathbf{r}_{\gamma'}(s)|^2},
\end{gathered}
\end{equation}    
\normalsize
where we use the classical identities \cite{Goldstein}
\begin{equation}
\begin{gathered}
\frac{\partial R^{\gamma}_{\mathcal{A}}}{\partial \mathbf{r}_i}=-\mathbf{P}^{i}_{\gamma}(\mathbf{r}_i,\mathbf{r}_f,t), \\
\frac{\partial R^{\gamma}_{\mathcal{A}}}{\partial \mathbf{r}_f}=\mathbf{P}^{f}_{\gamma}(\mathbf{r}_i,\mathbf{r}_f,t),
\end{gathered}
\end{equation}
for the initial and final momentum for the path $\gamma$, and similarly for the path $\gamma'$. 

The integral over $\mathbf{y}_i$ can now be performed to obtain
\small
\begin{equation}
\begin{aligned}
\mathcal{W}_{\mathcal{A}}\Big(\mathbf{r}_i,(\mathbf{P}^{i}_{\gamma}+\mathbf{P}^{i}_{\gamma'})/2\Big)= \frac{1}{(2\pi\hbar)^2}\int & d\mathbf{y}_i \rho_{\mathcal{A}}(\mathbf{r}_i+\mathbf{y}_i/2,\mathbf{r}_i-\mathbf{y}_i/2) \\& \mathrm{e}^{-\frac{i}{\hbar}(\mathbf{P}^{i}_{\gamma}+\mathbf{P}^{i}_{\gamma'})\cdot \mathbf{y}_i/2},
\end{aligned}
\end{equation}
\normalsize
where the initial Wigner function \cite{wigner} of the central system with initial momentum $(\mathbf{P}^{i}_{\gamma}+\mathbf{P}^{i}_{\gamma'})/2$ appears, to arrive at the expression
\begin{equation}\label{van_vleck_f}
\begin{gathered}
\rho^{\text{sc}}_{\mathcal{A}}(\mathbf{r}_f+\mathbf{y}_f/2,\mathbf{r}_f-\mathbf{y}_f/2,t)=\int_A d\mathbf{r}_i 
\sum_{\gamma: \mathbf{r}_i \to \mathbf{r}_f} \sum_{\gamma': \mathbf{r}_i \to \mathbf{r}_f} A_{\gamma}A^*_{\gamma'} \\ \mathcal{W}_{\mathcal{A}}\Big(\mathbf{r}_i,(\mathbf{P}^{i}_{\gamma}+\mathbf{P}^{i}_{\gamma'})/2\Big)  \mathrm{e}^{\frac{i}{\hbar}(R^{\gamma}_{\mathcal{A}}-R^{\gamma'}_{\mathcal{A}})} \\
\mathrm{e}^{\frac{i}{\hbar}(\mathbf{P}^{f}_{\gamma}+\mathbf{P}^{f}_{\gamma'})\cdot \mathbf{y}_f/2} 
\mathrm{e}^{-\alpha \int_{0}^{t} ds |\mathbf{r}_{\gamma}(s)-\mathbf{r}_{\gamma'}(s)|^2}.
\end{gathered}
\end{equation}  
  
A fully phase space representation is obtained after multiplying Eq.~(\ref{van_vleck_f}) by $\mathrm{e}^{-\frac{i}{\hbar}\mathbf{p}_f \cdot \mathbf{y}_f}$, and integrating over the variable $\mathbf{y}_f$. 
The left-hand side of Eq.~(\ref{van_vleck_f}) transforms then into the Wigner function of the central system at time $t$ and momentum $\mathbf{p}_f$, and the right-hand side gives just a delta function after the $\mathbf{y}_f$-integration. 

All together then, we obtain the important result for the time evolution of the Wigner function of the particle

 \begin{equation}\label{Wigner}
 \begin{gathered}
 \mathcal{W}^{\text{sc}}_{\mathcal{A}}(\mathbf{r}_f,\mathbf{p}_f,t)=\int_A d\mathbf{r}_i 
 \sum_{\gamma: \mathbf{r}_i \to \mathbf{r}_f} \sum_{\gamma': \mathbf{r}_i \to \mathbf{r}_f} A_{\gamma}A^*_{\gamma'} \\ \mathcal{W}_{\mathcal{A}}\Big(\mathbf{r}_i,(\mathbf{P}^{i}_{\gamma}+\mathbf{P}^{i}_{\gamma'})/2\Big)  \mathrm{e}^{\frac{i}{\hbar}(R^{\gamma}_{\mathcal{A}}-R^{\gamma'}_{\mathcal{A}})} \\
 \delta \Big(\mathbf{p}_f-(\mathbf{P}^{f}_{\gamma}+\mathbf{P}^{f}_{\gamma'})/2\Big)
  \; \mathrm{e}^{-\alpha \int_{0}^{t} ds |\mathbf{r}_{\gamma}(s)-\mathbf{r}_{\gamma'}(s)|^2},
 \end{gathered}
 \end{equation}   
involving a sum over pair of trajectories starting at point $\mathbf{r}_i$ and ending at $\mathbf{r}_f$ with the constraint in their final momentum. As mentioned before, the integration in Eq.~(\ref{Wigner}) runs over the area $A$ of the cavity, as appropriate for the calculation of expectation values of observables of the form $\hat{O} \chi_{A}(\hat{q})$ where $\chi_{A}(q)$ is the corresponding characteristic function.  

In the following subsections we will assume that we have introduced a local time average in Eq.~(\ref{Wigner}) in order to neglect highly oscillatory terms in the double sum.  

\subsection{Diagonal approximation} \label{diag}
Eq.~(\ref{Wigner}) represents the Wigner function, projected in a cavity of area $A$, of the central system at time $t$ in the semiclassical approximation, evolving from the initial Wigner function at time $t=0$. The most important contribution to the sum over pair of trajectories in (\ref{Wigner}) comes from the so-called diagonal approximation, where the two trajectories are identical $\gamma=\gamma'$.  In this case Eq.~(\ref{Wigner}) reads
 \begin{equation}\label{Wigner_diagonal}
\begin{gathered}
\mathcal{W}^{\text{dg}}_{\mathcal{A}}(\mathbf{r}_f,\mathbf{p}_f,t)=\int_A d\mathbf{r}_i 
\sum_{\gamma: \mathbf{r}_i \to \mathbf{r}_f} |A_{\gamma}|^2 \mathcal{W}_{\mathcal{A}}(\mathbf{r}_i, \mathbf{P}^{i}_{\gamma})  
\delta(\mathbf{p}_f-\mathbf{P}^{f}_{\gamma}),
\end{gathered}
\end{equation}     
where naturally the decoherence contribution has disappeared since it would involve off-diagonal terms.

It is important to note that, for a system constrained to be in a closed area, we can use the amplitude $|A_{\gamma}|^2 =\rm{det}\Big|\frac{\partial \mathbf{P}^{f}_{\gamma} }{\partial \mathbf{r}_i}\Big|$, as a Jacobian transformation from initial position to final momentum, to get
\small
\begin{equation}\label{diagomal_1}
\begin{aligned}
\mathcal{W}^{\text{dg}}_{\mathcal{A}}(\mathbf{r}_f,\mathbf{p}_f,t)= &\int d\mathbf{P}_f \delta(\mathbf{p}_f-\mathbf{P}_f)  \mathcal{W}_{\mathcal{A}}\Big(\mathbf{r}_i(\mathbf{r}_f,\mathbf{P}_f,t), \mathbf{p}_i(\mathbf{r}_f,\mathbf{P}_f,t)\Big) \\ = &  \; \mathcal{W}_{\mathcal{A}}\Big(\mathbf{r}_i(\mathbf{r}_f,\mathbf{p}_f,t), \mathbf{p}_i(\mathbf{r}_f,\mathbf{p}_f,t)\Big),
\end{aligned}
\end{equation}
\normalsize
which says that the Wigner function at time $t$ is simply obtained in terms of the initial Wigner function by  rigidly transporting backwards its values along the solution of the classical equations of motion $(\mathbf{r}_f,\mathbf{p}_f)=(\mathbf{r}_{f}(\mathbf{r}_i,\mathbf{p}_i,t),\mathbf{p}_{f}(\mathbf{r}_i,\mathbf{p}_i,t))$.
This is the so-called Truncated Wigner approximation \cite{PhysRevA-polk,POLKOVNIKOV,Truncated_W,PhysRevA-Drummond}, expressing in the semiclassical limit the evolution of quantum mechanical states by means of classical evolution of the corresponding Wigner function.

In the case of interest here, however, we project the Wigner function in a cavity and thus Eq.~(\ref{Wigner_diagonal}) gives the diagonal approximation of the projected Wigner function, which allows us to calculate local observables inside the cavity.  

Using the sum rule for open systems \cite{Richter2002},   
and assuming a state with a well-defined mean energy $E_0$, we get
\small
\begin{equation}\label{diagomal_open}
\begin{gathered}
\mathcal{W}^{\text{dg}}_{\mathcal{A}}(\mathbf{r}_f,\mathbf{p}_f,t)=\int_A d\mathbf{P}_f \delta(\mathbf{p}_f-\mathbf{P}_f) \mathrm{e}^{-t/\tau_D} \\ \mathcal{W}_{\mathcal{A}}\Big(\mathbf{r}_i(\mathbf{r}_f,\mathbf{P}_f,t), \mathbf{p}_i(\mathbf{r}_f,\mathbf{P}_f,t)\Big) \\ =  \mathrm{e}^{-t/\tau_D} \mathcal{W}_{\mathcal{A}}\Big(\mathbf{r}_i(\mathbf{r}_f,\mathbf{p}_f,t), \mathbf{p}_i(\mathbf{r}_f,\mathbf{p}_f,t)\Big),
\end{gathered}
\end{equation}   
\normalsize
where $1/\tau_D$ is the classical escape rate at energy $E_0$. Eq.~(\ref{diagomal_open}) results in an exponential decay of the projected Wigner function inside the cavity. In particular, the probability to find the particle inside the cavity at time $t$ can be obtained as $\int_A d\mathbf{r}_f  d\mathbf{p}_f \mathcal{W}^{\text{dg}}_{\mathcal{A}}(\mathbf{r}_f,\mathbf{p}_f,t)$, and gives the result for the classical survival probability in \cite{Gutierrezprl}.

While in the diagonal approximation the decohrence factor in Eq.~(\ref{Wigner}), $-\alpha \int_{0}^{t} ds |\mathbf{r}_{\gamma}(s)-\mathbf{r}_{\gamma'}(s)|^2$, cancels  out, the leading-order quantum correction to (\ref{Wigner}) for a chaotic system, the so-called loop contributions,
involves pairs of correlated trajectories which are not identical all the time and thus could reveal interference effects between the involved paths. This is the topic of the next subsection.  

\subsection{Loop corrections} \label{loop}

\begin{figure}
\centering
\includegraphics[width=0.4\textwidth]{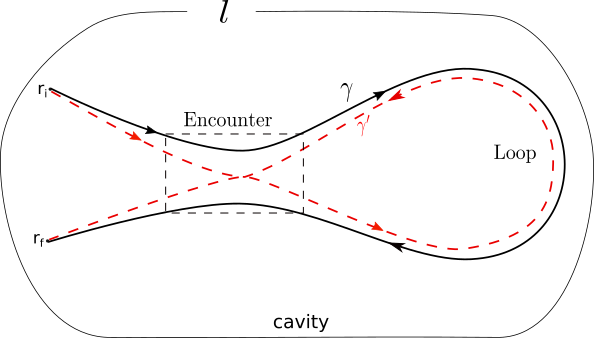}
\caption{A typical pair of correlated trajectories $\gamma$, $\gamma'$ inside the cavity. The trajectories  differ from each other inside the encounter   region where they change partners, but remain close, and after leaving the encounter   they form a loop, one trajectory following the time-reversed path of the other. The draw is in configuration space and the arrows show the direction of the momentum. This is an example of a $2$-leg diagram where the encounter   is fully developed between the endpoints. }
\label{fig1}
\end{figure}

The leading order quantum correction to the time evolution of the projected Wigner function in Eq.~(\ref{Wigner}) comes from pair of trajectories $\gamma, \gamma'$ which are identical to each other except in a so-called self-encounter region \cite{Sieber2001}, where they remain close to each other but shift partners, as shown in Fig.~\ref{fig1} (for a system with time reversal symmetry). 

In this scenario there are three diagrams whose contributions have to be added within the leading-order loop correction: when the encounter takes place at the beginning (or at the end) of the trajectory, called $1$-leg-loops, and when the encounter is fully developed in the region between the endpoints of the trajectory, called $2$-leg-loops.    

\begin{figure}
\centering
\includegraphics[width=0.4\textwidth]{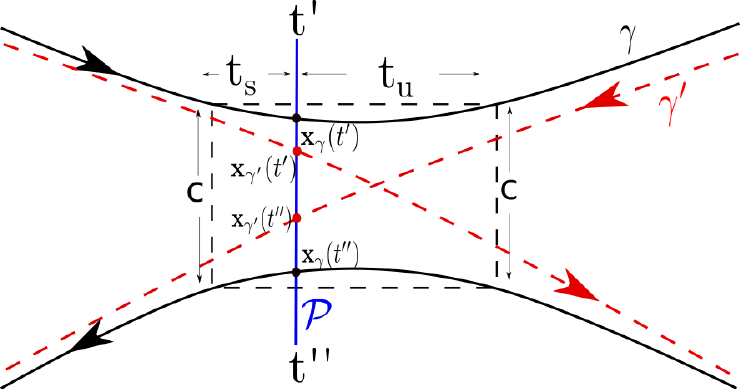}
\caption{A Poincare surface of section $\mathcal{P}$ is placed inside the encounter, and there we select a reference point in phase-space $\mathbf{x}_{\gamma}(t')$ of the trajectory $\gamma$, when it first reaches $\mathcal{P}$ at time $t'$. A local reference frame is constructed at this point with $(s,u)$ the local stable and unstable manifold. Using this frame, the trajectory $\gamma'$ can be described within the linerized regime inside the encounter. }
\label{fig2}
\end{figure}

Let us sketch the calculation for the contribution of the $2$-leg diagram. 
As usual \cite{Libro_chaos} we place a Poincare surface of section $\mathcal{P}$ at any point inside the encounter, as shown in Fig.~\ref{fig2}. The trajectory $\gamma$ first reaches $\mathcal{P}$ at time $t'$ and then, after leaving the encounter, forms a loop and returns back to the encounter, reaching again $\mathcal{P}$ a second time at $t''$.

Inside the encounter the two trajectories $\gamma, \gamma'$ are different form each other, but remain close. In this region we can describe one trajectory in terms of a local coordinate system localized on the other trajectory. As shown in Fig.~\ref{fig2} we select a reference point in phase-space $\mathbf{x}_{\gamma}$ at time $t'$ and construct a local coordinate system $(s,u)$ based on the stable and unstable local manifold, where $t_u$ is the time trajectory $\gamma'$ needs to leave the encounter (to escape the linearized regime) in the unstable direction, and in a similar way is defined $t_s$ along the stable direction. In this way the whole trajectory is divided into four parts: three links and the encounter region.

In the initial and final link $\gamma$ and $\gamma'$ are identical and then the action difference and the decoherence term from Eq.~(\ref{Wigner}) vanish. In the second link, there is a loop in which $\gamma$ and $\gamma'$, after leaving the encounter region, have shifted partners and then one trajectory follows the time-reversed path of the other. The time the first stretch needs to travel the encounter is  $t_{enc}=t_u+t_s$, the duration of the loop is called $t_{\text{loop}}$, and thus the second time $\gamma$ reaches $\mathcal{P}$ is $t''=t'+2t_u+t_{\text{loop}}$.  

The key observation is that the decoherence term inside the loop is no longer zero. Moreover, since being $\gamma$ the time reversed of $\gamma'$, the paths $\mathbf{r}_{\gamma}(s)$ and $\mathbf{r}_{\gamma'}(s)$ can be treated as uncorrelated through the loop. The important role of this type of non-diagonal suppression has been studied in the framework of closed systems by calculating its effect on the loss of purity \cite{Pitetjean}.

With this division of the trajectory the decoherence term itself is splits into
\small
\begin{equation} \label{difference}
\begin{gathered}
    \int_{0}^{t} |\mathbf{r}_{\gamma}(\tau)-\mathbf{r}_{\gamma'}(\tau)|^2=\int_{0}^{t'-t_s} |\mathbf{r}_{\gamma}(\tau)-\mathbf{r}_{\gamma'}(\tau)|^2 \\ +\int_{t'-t_s}^{t'+t_u} |\mathbf{r}_{\gamma}(\tau)-\mathbf{r}_{\gamma'}(\tau)|^2+ \int_{t'+t_u}^{t'+t_u+t_{\text{loop}}} |\mathbf{r}_{\gamma}(\tau)-\mathbf{r}_{\gamma'}(\tau)|^2 \\ + \int_{t'+t_u+t_{\text{loop}}}^{t'+2t_u+t_s+t_{\text{loop}}} |\mathbf{r}_{\gamma}(\tau)-\mathbf{r}_{\gamma'}(\tau)|^2 \\ + \int_{t'+2t_u+t_s+t_{\text{loop}}}^{t} |\mathbf{r}_{\gamma}(\tau)-\mathbf{r}_{\gamma'}(\tau)|^2,
\end{gathered}
\end{equation}
\normalsize
where the first and last integrals represent the first and third link respectively. There $\gamma$ and $\gamma'$ are identical and thus the integrals vanish. 
\\

\textit{Inside the loop}. The important contribution to decoherence comes from pair of trajectories inside the loop. To calculate this contribution we apply ergodic arguments: due to the chaotic nature of the system we transform the time integral of the squared difference in Eq.~(\ref{difference}) into a variance of position $\sigma^2$. To this end we add to the integrand the phase-space average position value  $\braket{\mathbf{r}}_{E_{\gamma}}$, where $E_{\gamma}$ denotes the energy of the trajectory $\gamma$, which is the same energy of $\gamma'$, and write    
\small
\begin{equation}
\begin{aligned}
 &\int_{t'+t_u}^{t'+t_u+t_{\text{loop}}}  d\tau  |\mathbf{r}_{\gamma}(\tau)-\braket{\mathbf{r}}_{E_{\gamma}}+
 \braket{\mathbf{r}}_{E_{\gamma}}-\mathbf{r}_{\gamma'}(\tau)|^2 \\ = & \int_{t'+t_u}^{t'+t_u+t_{\text{loop}}} d\tau \;\Big( |\mathbf{r}_{\gamma}(\tau)-\braket{\mathbf{r}}_{E_{\gamma}}|^2+|\mathbf{r}_{\gamma'}(\tau)-\braket{\mathbf{r}}_{E_{\gamma}}|^2 \\ &+2(\mathbf{r}_{\gamma'}(\tau)-\braket{\mathbf{r}}_{E_{\gamma}})^T\cdot (\mathbf{r}_{\gamma'}(\tau)-\braket{\mathbf{r}}_{E_{\gamma}})\Big) \\
= & \; 2 t_{\text{loop}} \braket{(\mathbf{r}-\braket{\mathbf{r}}_{E_{\gamma}})^2} :=\; 2 t_{\text{loop}} \sigma^2. \noindent
\end{aligned}
\end{equation} 
\normalsize

To obtain the last line we use the relation
 \begin{equation}
 \frac{1}{T}\int_{0}^{T}d\tau f(\mathbf{r}_{\gamma}(\tau),\mathbf{p}_{\gamma}(\tau))=\braket{f(\mathbf{r},\mathbf{p})}_{E_{\gamma}},
 \end{equation} 
to change the time integral into phase-space average $\braket{f}_{E_{\gamma}}$, with $\braket{\mathbf{r}-\braket{\mathbf{r}}_{E_{\gamma}}}=0$.
\newline 
 
\textit{Inside the encounter.} When the pair $(\gamma,\gamma')$ traverses the encounter   for the first time, that is in the time interval $[t'-t_s,t'+t_u]$, the difference $\mathbf{r}_{\gamma}(\tau)-\mathbf{r}_{\gamma'}(\tau)$ at any time $\tau$ within the interval, calculated from the reference point $\mathbf{x}_{\gamma}$, is given in the linearized regime by $\mathbf{r}_{\gamma}(\tau)-\mathbf{r}_{\gamma'}(\tau)=-u\mathrm{e}^{\lambda(\tau-t')}\tilde{e}_u(\mathbf{x}_{\gamma}(\tau))$. Where $u$ ($s$) is the coordinate in the unstable (stable) manifold, $\lambda$ is the Lyapunov exponent (assumed uniform) and $\tilde{e}_u(\mathbf{x}_{\gamma}(\tau))$ is a local unit vector pointing in the unstable direction at time $\tau$.

With this considerations the decoherence term inside the encounter during the first time interval can be evaluated to give
\small
\begin{equation}
 \int_{t'-t_s}^{t'+t_u} d\tau |\mathbf{r}_{\gamma}(s)-\mathbf{r}_{\gamma'}(\tau)|^2 = u^2  \int_{t'-t_s}^{t'+t_u} d\tau \: \mathrm{e}^{2\lambda(\tau-t')} |\tilde{e}_u(\mathbf{x}_{\gamma}(\tau))|^2, 
\end{equation}
\normalsize
where, in the semiclassical limit the precise time-dependence of $\tilde{e}_u(\mathbf{x}_{\gamma}(\tau))$ is effectively averaged over the phase space in order to take it out of the time integral as a constant $\eta$, whose exact value will not play any role in the final result.

In this way the last equation gives 
\begin{equation}
\int_{t'-t_s}^{t'+t_u} d\tau |\mathbf{r}_{\gamma}(s)-\mathbf{r}_{\gamma'}(\tau)|^2= \eta \frac{c^2}{2\lambda} \Big(1-\Big(\frac{su}{c^2}\Big)^2\Big),
\end{equation} 
where the factor $c$ is a classical scale constant characterizing the linearized regime. A similar calculation is carried out for the second time interval inside the encounter, and we obtain finally the total contribution of the decoherence term, 
\begin{equation}\label{decoh_total}
\int_{0}^{t} d\tau |\mathbf{r}_{\gamma}(s)-\mathbf{r}_{\gamma'}(\tau)|^2= \eta \frac{c^2}{\lambda} \Big(1-\Big(\frac{su}{c^2}\Big)^2\Big)+2t_{\text{loop}} \sigma^2,
\end{equation}
\newline
while it can be shown \cite{Libro_chaos} that the action difference results in $R^{\gamma}_{\mathcal{A}}-R^{\gamma'}_{\mathcal{A}}=su$.  
The final ingredient to calculate the loop correction to Eq.~(\ref{Wigner}) is the density of trajectories with a self-encounter $\omega_{\gamma}(s,u,t',t_{\text{loop}})$ with action difference $su$ and loop duration $t_{\text{loop}}$. Using the mixing property of chaotic systems, this density can be approximated by $\omega_{\gamma}(s,u,t',t_{\text{loop}})=\frac{1}{\Omega_{\gamma}t_{\rm{enc}}(s,u)}$, with $\Omega_{\gamma}$ the phase-space volume \cite{Daniel}.  In this way, we have characterized all the partner orbits $\gamma'$ for a given $\gamma$ trajectory, and we can perform the sum over $\gamma$ by taking $|A_{\gamma}|^2$ in Eq.~(\ref{Wigner}) as a Jacobian transformation, using the sum rule as in section \ref{diag}.

Finally, as shown in \cite{Gutierrezprl}, we take into account that the quantum survival probability is augmented by the factor $\mathrm{e}^{t_{\rm{enc}}/\tau_D}$, and using Eq.~(\ref{decoh_total}), the first quantum correction to Eq.~(\ref{Wigner}) is given by
\begin{equation}\label{loops}
\begin{gathered}
 \mathcal{W}^{\text{loop}}_{\mathcal{A}}(\mathbf{r}_f,\mathbf{p}_f,t)_{\rm{2-legs}} = \mathcal{W}_{\mathcal{A}}\Big(\mathbf{r}_i(\mathbf{r}_f,\mathbf{p}_f,t), \mathbf{p}_i(\mathbf{r}_f,\mathbf{p}_f,t)\Big) \\ \int_{-c^2}^{c^2}dsdu   \int_{t_s}^{t-2t_u-t_s}dt'\int_{0}^{t-t'-2t_u-t_s}dt_{\text{loop}} \\ 
 \times \frac{\mathrm{e}^{-(t-t_{\rm{enc}}(s,u))}}{\Omega t_{\text{enc}}(s,u)}\mathrm{e}^{-\alpha\eta \frac{c^2}{\lambda} (1-(\frac{su}{c^2})^2)} \mathrm{e}^{-\alpha2t_{\text{loop}} \sigma^2},
 \end{gathered}
\end{equation}
where the limits of the integration reflect the fact that we need a minimum time $t_u+t_s$ to form an encounter region, and the variables $(s,u)$ can not grow beyond the limit $c$. The encounter time reads $t_{\rm{enc}}=\lambda^{-1}\log(c^2/|su|)$.

An integral similar to Eq.~(\ref{loops}) is obtained for the contribution of the $1$-leg diagrams, but the time intervals for $t'$ and $t_{\text{loop}}$ have to be adjusted to account for the fact that encounters at the beginning or at the end of the trajectory do not have time to fully develop. 

We evaluate the integral in (\ref{loops}), and the one coming from $1$-leg diagrams, in the semiclassical regime where $\lambda \tau_D$, $c^2/\hbar$ $\to \infty$, while $\alpha/\lambda \to 0$, to get

\begin{equation}\label{result}
\begin{gathered}
 \mathcal{W}^{\text{loop}}_{\mathcal{A}}(\mathbf{r}_f,\mathbf{p}_f,t)  =  \mathcal{W}_{\mathcal{A}}\Big(\mathbf{r}_i(\mathbf{r}_f,\mathbf{p}_f,t), \mathbf{p}_i(\mathbf{r}_f,\mathbf{p}_f,t)\Big) \\ 
\Bigg[ \frac{\tau_d^2}{T_H \tau_D}\mathrm{e}^{-t/\tau_D}\Big(
 \mathrm{e}^{-t/\tau_d}-1 \Big) 
 +\frac{\tau_d}{T_H \tau_D}t \; \mathrm{e}^{-t/\tau_D} \Bigg],
\end{gathered}    
\end{equation}
where we have defined the \textit{decoherence time}, 
\begin{equation}
\tau_d=(2\alpha \sigma^2)^{-1},
\end{equation}
with the variance $\sigma^2$ giving an estimate of the average separation in position of two  correlated trajectories. 

\begin{figure} 
\centering
\includegraphics[width=0.4\textwidth]{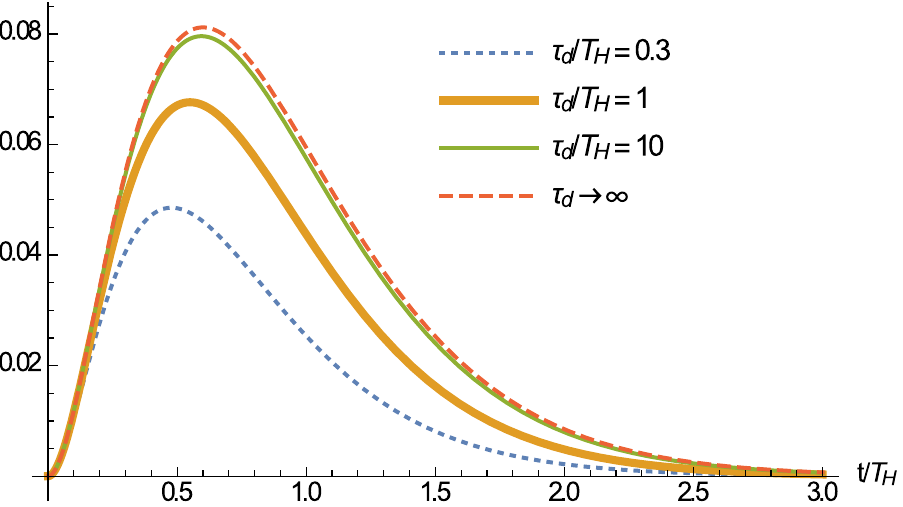}
\caption{Plot of the first quantum correction to Eq.~(\ref{diagomal_open}) (the function in brackets in Eq.~(\ref{result})) as a function of $t/T_H$ for a value $\tau_D/T_H=0.3$, and different $\tau_d/T_H$. The graphic shows a detriment of the first quantum correction to the survival probability, compared to the vanishing coupling result, due to the coupling of the system to the environment. The red (dashed) line, $\tau_d \to \infty$, represents the correction for vanishing coupling. }
\label{decohresult}
\end{figure}

Equation~(\ref{result}) is our main result.  As shown in Fig.~\ref{decohresult}, it gives an analytical result for the interplay between the quantum enhancement due to coherent interference effects coming from correlated trajectories inside the encounter region \cite{Gutierrezprl}, illustrated in red (dashed) line, and on the other hand, the detriment of the quantum survival probability, compared with the vanishing coupling result, due to decoherence effects depending on the temperature and the coupling strength, coming from uncorrelated trajectories inside the loop, which give rise to the term $\mathrm{e}^{-2\alpha \sigma^2 t}$. 

In the short-time regime, obtained by expanding $\mathrm{e}^{-t/\tau_d}$ for small $t/\tau_d$, Eq.~(\ref{result}) reads
\begin{equation}\label{result_2}
\begin{gathered}
  \mathcal{W}^{\text{loop}}_{\mathcal{A}}(\mathbf{r}_f,\mathbf{p}_f,t) = \mathcal{W}_{\mathcal{A}}\Big(\mathbf{r}_i(\mathbf{r}_f,\mathbf{p}_f,t), \mathbf{p}_i(\mathbf{r}_f,\mathbf{p}_f,t\Big) \\ \mathrm{e}^{-t/\tau_D}\Bigg[\frac{t^2}{2T_H\tau_D}-\frac{t^3}{6T_H\tau_D\tau_d}+\mathcal{O}(t^4/(\tau_d^2T_H\tau_D))\Bigg], 
\end{gathered}
\end{equation}
and we identify in the quadratic time-dependence the well-known result for the first quantum correction to the survival probability found in \cite{Gutierrezprl}, (see Eq.~(\ref{qm_survival})).

It is important to observe that when we close the opening of the cavity, $\tau_D \to \infty$, the loop contribution $ \mathcal{W}^{\text{loop}}_{\mathcal{A}}(\mathbf{r}_f,\mathbf{p}_f,t)$ in Eq.~(\ref{result}) vanishes. So in the closed-opening scenario, and when the system is only coupled to a bath which produces decoherence in position, all quantum loop corrections cancel out in the semiclassical limit.   

This cancellation of quantum loop corrections for a closed system with classical chaotic dynamics points to an extremely robust character of the diagonal approximation (and of the Truncated Wigner method), and can be understood as a generalization of the very nontrivial loop cancellation order by order in $\hbar$ shown in \cite{Kuipers_2009} for the integrated probability, where it simply accounts for unitarity of quantum evolution. The fact that loop corrections to the more fundamental (non-integrated) Wigner function  as we obtained manifest only when the system is open is indeed a fascinating observation for which a clear physical mechanism is still not at hand.  

\textit{Ehrenfest-time effects}. As a final stage we calculate explicitly the dependence of our result with the Ehrenfest-time. This is the time scale above which quantum interference becomes important and is generally defined as $t_E=\lambda^{-1}\log(c^2/\hbar)$ \cite{Jacquod2006}. 

\begin{figure}
\centering
\includegraphics[width=0.4\textwidth]{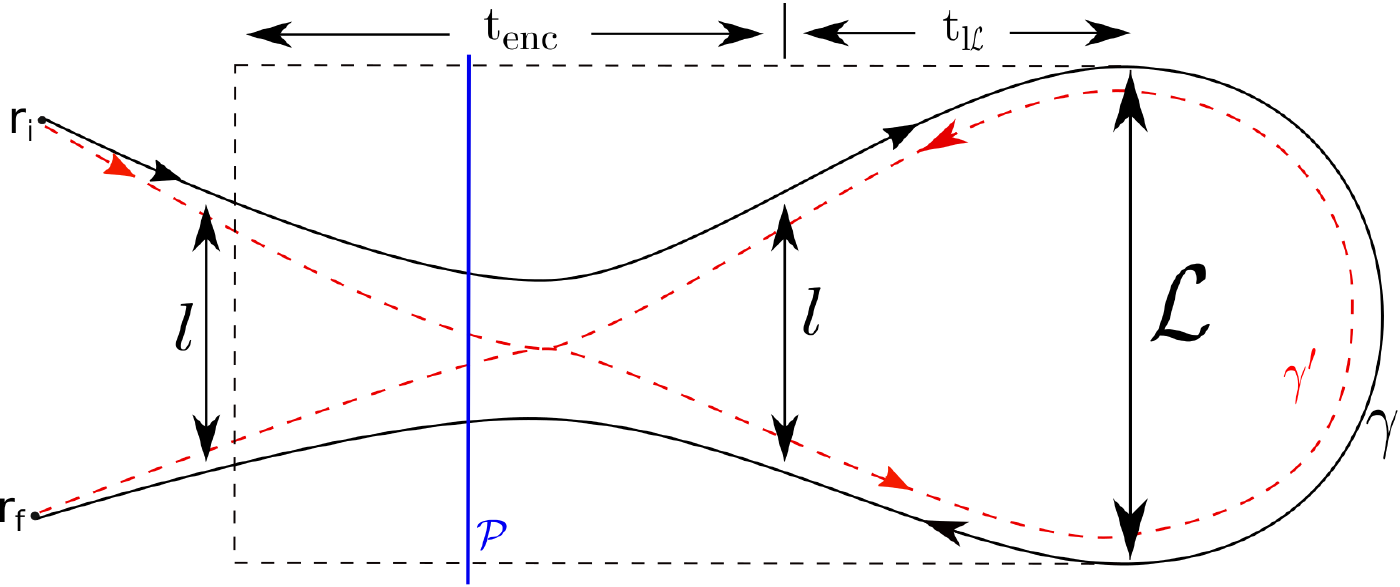}
\caption{In the case of non-vanishing Ehrenfest time, when the stretches escape the encounter they need to be separated a distance of the order of the cavity opening size $l$ in order for them to leave the encounter region in a uncorrelated manner. And to form a loop the stretches have to be separated a distance of the order of the size of the cavity $\mathcal{L}$.}
\label{figehrenfest}
\end{figure}

So we impose the total time of the trajectory to be longer than $2t_E$ to allow the formation of a minimal encounter region.
On the other hand, as shown in \cite{Gutierrez2009}, for a cavity with opening size $l$ we require the encounter stretches to escape the encounter when their separation is of the order $l$, in order to them leave the encounter in an uncorrelated manner.
Moreover, as shown in Fig.~\ref{figehrenfest}, 
on the right-hand side of the encounter the stretches should be separated a distance of the order of the size of he cavity $\mathcal{L}$, in order to close themselves forming a loop. This imposes a minimum time of the loop, which is $2t_{l\mathcal{L}}$, where $t_{l\mathcal{L}}=\lambda^{-1}\log(\mathcal{L}/l)$.  
Indeed, with this considerations it is clear that the variance in position should be of the order of the size of the cavity, giving $\tau_d=(2\alpha \mathcal{L}^2)^{-1}$.

With these restrictions, and redefining appropriately the time limits, we solve the integrals in Eq.~(\ref{loops}), introducing a step function $\theta(t-2t_E-2t_{l\mathcal{L}})$, and similarly for the $1$-leg diagrams, to get
\begin{equation}\label{result_ehrenfest}
\begin{aligned}
 & \mathcal{W}^{\text{loop}}_{\mathcal{A}}(\mathbf{r}_f,\mathbf{p}_f,t)  =  \mathcal{W}_{\mathcal{A}}\Big(\mathbf{r}_i(\mathbf{r}_f,\mathbf{p}_f,t), \mathbf{p}_i(\mathbf{r}_f,\mathbf{p}_f,t)\Big) \\ 
& \Bigg[ \frac{\tau_d^2}{T_H \tau_D}\mathrm{e}^{-(t-t_E)/\tau_D}\Big(
 \mathrm{e}^{-(t-2t_E)/\tau_d}-\mathrm{e}^{-2t_{l\mathcal{L}}/\tau_d} \Big) \\ 
 & +\frac{(t-2t_E-2t_{l\mathcal{L}})\tau_d}{T_H \tau_D}\mathrm{e}^{-(t-t_E)/\tau_D}\mathrm{e}^{-2t_{l\mathcal{L}}/\tau_d}\Bigg]\theta(t-2t_E-2t_{l\mathcal{L}}), 
\end{aligned}
\end{equation}
that completes the full semiclassical analysis.

\section{Conclusions} \label{conlc}

In this paper, we provide a complete picture of the effect of decoherence on the coherent quantum corrections to classical population decay in chaotic cavities. It begins with the construction of the semiclassical Wigner representation of a chaotic particle weakly coupled to an environment within the Caldeira-Legget model. This representation Eq.~(\ref{Wigner}) consists of a double sum of classical trajectories, and we show that it is the difference between these pair of trajectories that generates an exponential decay in the Wigner function due to positional decoherence. Coherent effects due to path interference are made explicit when projecting this Wigner function in an open cavity of area $A$, appropriate to calculate local observables inside the cavity. We find the first-order quantum correction due to path interference to the time evolution, which leads to a universal non-monotonous form depending on the properties of the cavity and the  bath-coupling parameters. In particular, the interplay between a coherent enhancement of the survival probability, coming from correlated trajectories inside an encounter region, and on the other hand, the decoherence effect coming from uncorrelated trajectories inside a loop, produces a detriment of the quantum survival probability compared to the scenario of vanishing coupling. Our analysis is completed by calculating the explicit dependence of this first-order quantum correction with the Ehrenfest-time. 
\\

\textbf{Acknowledgments}. C.M. acknowledges financial support from the German Academic Exchange Service DAAD. We would like to thank Klaus Richter for illuminating discussions.   
\bibliography{bib}

\begin{thebibliography}{31}%
\makeatletter
\providecommand \@ifxundefined [1]{%
 \@ifx{#1\undefined}
}%
\providecommand \@ifnum [1]{%
 \ifnum #1\expandafter \@firstoftwo
 \else \expandafter \@secondoftwo
 \fi
}%
\providecommand \@ifx [1]{%
 \ifx #1\expandafter \@firstoftwo
 \else \expandafter \@secondoftwo
 \fi
}%
\providecommand \natexlab [1]{#1}%
\providecommand \enquote  [1]{``#1''}%
\providecommand \bibnamefont  [1]{#1}%
\providecommand \bibfnamefont [1]{#1}%
\providecommand \citenamefont [1]{#1}%
\providecommand \href@noop [0]{\@secondoftwo}%
\providecommand \href [0]{\begingroup \@sanitize@url \@href}%
\providecommand \@href[1]{\@@startlink{#1}\@@href}%
\providecommand \@@href[1]{\endgroup#1\@@endlink}%
\providecommand \@sanitize@url [0]{\catcode `\\12\catcode `\$12\catcode
  `\&12\catcode `\#12\catcode `\^12\catcode `\_12\catcode `\%12\relax}%
\providecommand \@@startlink[1]{}%
\providecommand \@@endlink[0]{}%
\providecommand \url  [0]{\begingroup\@sanitize@url \@url }%
\providecommand \@url [1]{\endgroup\@href {#1}{\urlprefix }}%
\providecommand \urlprefix  [0]{URL }%
\providecommand \Eprint [0]{\href }%
\providecommand \doibase [0]{http://dx.doi.org/}%
\providecommand \selectlanguage [0]{\@gobble}%
\providecommand \bibinfo  [0]{\@secondoftwo}%
\providecommand \bibfield  [0]{\@secondoftwo}%
\providecommand \translation [1]{[#1]}%
\providecommand \BibitemOpen [0]{}%
\providecommand \bibitemStop [0]{}%
\providecommand \bibitemNoStop [0]{.\EOS\space}%
\providecommand \EOS [0]{\spacefactor3000\relax}%
\providecommand \BibitemShut  [1]{\csname bibitem#1\endcsname}%
\let\auto@bib@innerbib\@empty
\bibitem [{\citenamefont {Zurek}(2003)}]{Zurek2003}%
  \BibitemOpen
  \bibfield  {author} {\bibinfo {author} {\bibfnamefont {W.~H.}\ \bibnamefont
  {Zurek}},\ }\href {\doibase 10.1103/RevModPhys.75.715} {\bibfield  {journal}
  {\bibinfo  {journal} {Reviews of Modern Physics}\ }\textbf {\bibinfo {volume}
  {75}},\ \bibinfo {pages} {715} (\bibinfo {year} {2003})}\BibitemShut
  {NoStop}%
\bibitem [{\citenamefont {Goldstein}(2002)}]{Goldstein}%
  \BibitemOpen
  \bibfield  {author} {\bibinfo {author} {\bibfnamefont {H.}~\bibnamefont
  {Goldstein}},\ }\href@noop {} {\emph {\bibinfo {title} {{Classical
  Mechanics}}}}\ (\bibinfo  {publisher} {Addison-Wesley},\ \bibinfo {year}
  {2002})\BibitemShut {NoStop}%
\bibitem [{\citenamefont {Schlosshauer}(2007)}]{Schlosshauer}%
  \BibitemOpen
  \bibfield  {author} {\bibinfo {author} {\bibfnamefont {M.}~\bibnamefont
  {Schlosshauer}},\ }\href@noop {} {\emph {\bibinfo {title} {{Decoherence and
  theQuantum-To-Classical Transition}}}},\ edited by\ \bibinfo {editor}
  {\bibfnamefont {F.~C.}\ \bibnamefont {Springer}}\ (\bibinfo {year}
  {2007})\BibitemShut {NoStop}%
\bibitem [{\citenamefont {von Neumann}(1932)}]{Neumann}%
  \BibitemOpen
  \bibfield  {author} {\bibinfo {author} {\bibfnamefont {J.}~\bibnamefont {von
  Neumann}},\ }\bibfield  {booktitle} {\emph {\bibinfo {booktitle}
  {{Mathematische Grundlagen der Quantenmeckanik}}},\ }\href@noop {} {\
  (\bibinfo {year} {1932})}\BibitemShut {NoStop}%
\bibitem [{\citenamefont {Zurek}(1981)}]{Zurek1981}%
  \BibitemOpen
  \bibfield  {author} {\bibinfo {author} {\bibfnamefont {W.~H.}\ \bibnamefont
  {Zurek}},\ }\href {\doibase 10.1103/PhysRevD.24.1516} {\bibfield  {journal}
  {\bibinfo  {journal} {Physical Review D}\ }\textbf {\bibinfo {volume} {24}},\
  \bibinfo {pages} {1516} (\bibinfo {year} {1981})}\BibitemShut {NoStop}%
\bibitem [{\citenamefont {Zurek}(1982)}]{Zurek1982}%
  \BibitemOpen
  \bibfield  {author} {\bibinfo {author} {\bibfnamefont {W.~H.}\ \bibnamefont
  {Zurek}},\ }\href {\doibase 10.1103/PhysRevD.26.1862} {\bibfield  {journal}
  {\bibinfo  {journal} {Physical Review D}\ }\textbf {\bibinfo {volume} {26}},\
  \bibinfo {pages} {1862} (\bibinfo {year} {1982})}\BibitemShut {NoStop}%
\bibitem [{\citenamefont {Raimond}\ \emph {et~al.}(2001)\citenamefont
  {Raimond}, \citenamefont {Brune},\ and\ \citenamefont
  {Haroche}}]{Experiment1}%
  \BibitemOpen
  \bibfield  {author} {\bibinfo {author} {\bibfnamefont {J.~M.}\ \bibnamefont
  {Raimond}}, \bibinfo {author} {\bibfnamefont {M.}~\bibnamefont {Brune}}, \
  and\ \bibinfo {author} {\bibfnamefont {S.}~\bibnamefont {Haroche}},\ }\href
  {\doibase 10.1103/RevModPhys.73.565} {\bibfield  {journal} {\bibinfo
  {journal} {Reviews of Modern Physics}\ }\textbf {\bibinfo {volume} {73}},\
  \bibinfo {pages} {565} (\bibinfo {year} {2001})}\BibitemShut {NoStop}%
\bibitem [{\citenamefont {Kim}\ \emph {et~al.}(1999)\citenamefont {Kim},
  \citenamefont {{Fonseca Romero}}, \citenamefont {Horiguti}, \citenamefont
  {Davidovich}, \citenamefont {Nemes},\ and\ \citenamefont {{de Toledo
  Piza}}}]{Experiment2}%
  \BibitemOpen
  \bibfield  {author} {\bibinfo {author} {\bibfnamefont {J.~I.}\ \bibnamefont
  {Kim}}, \bibinfo {author} {\bibfnamefont {K.~M.}\ \bibnamefont {{Fonseca
  Romero}}}, \bibinfo {author} {\bibfnamefont {A.~M.}\ \bibnamefont
  {Horiguti}}, \bibinfo {author} {\bibfnamefont {L.}~\bibnamefont
  {Davidovich}}, \bibinfo {author} {\bibfnamefont {M.~C.}\ \bibnamefont
  {Nemes}}, \ and\ \bibinfo {author} {\bibfnamefont {A.~F.~R.}\ \bibnamefont
  {{de Toledo Piza}}},\ }\href {\doibase 10.1103/PhysRevLett.82.4737}
  {\bibfield  {journal} {\bibinfo  {journal} {Physical Review Letters}\
  }\textbf {\bibinfo {volume} {82}},\ \bibinfo {pages} {4737} (\bibinfo {year}
  {1999})}\BibitemShut {NoStop}%
\bibitem [{\citenamefont {{Hackerm{\"u}ller L.}}\ \emph
  {et~al.}(2003)\citenamefont {{Hackerm{\"u}ller L.}}, \citenamefont
  {{Hornberger K.}}, \citenamefont {{Brezger B.}}, \citenamefont {{Zeilinger
  A.}},\ and\ \citenamefont {{Arndt M.}}}]{Expetiment3}%
  \BibitemOpen
  \bibfield  {author} {\bibinfo {author} {\bibnamefont {{Hackerm{\"u}ller
  L.}}}, \bibinfo {author} {\bibnamefont {{Hornberger K.}}}, \bibinfo {author}
  {\bibnamefont {{Brezger B.}}}, \bibinfo {author} {\bibnamefont {{Zeilinger
  A.}}}, \ and\ \bibinfo {author} {\bibnamefont {{Arndt M.}}},\ }\href
  {\doibase 10.1007/s00340-003-1312-6} {\bibfield  {journal} {\bibinfo
  {journal} {Applied Physics B}\ }\textbf {\bibinfo {volume} {77}},\ \bibinfo
  {pages} {781} (\bibinfo {year} {2003})}\BibitemShut {NoStop}%
\bibitem [{\citenamefont {{Hackerm{\"u}ller Lucia}}\ \emph
  {et~al.}(2004)\citenamefont {{Hackerm{\"u}ller Lucia}}, \citenamefont
  {{Hornberger Klaus}}, \citenamefont {{Brezger Bj{\"o}rn}}, \citenamefont
  {{Zeilinger Anton}},\ and\ \citenamefont {{Arndt Markus}}}]{Experiment4}%
  \BibitemOpen
  \bibfield  {author} {\bibinfo {author} {\bibnamefont {{Hackerm{\"u}ller
  Lucia}}}, \bibinfo {author} {\bibnamefont {{Hornberger Klaus}}}, \bibinfo
  {author} {\bibnamefont {{Brezger Bj{\"o}rn}}}, \bibinfo {author}
  {\bibnamefont {{Zeilinger Anton}}}, \ and\ \bibinfo {author} {\bibnamefont
  {{Arndt Markus}}},\ }\href {\doibase 10.1038/nature02276} {\bibfield
  {journal} {\bibinfo  {journal} {Nature}\ }\textbf {\bibinfo {volume} {427}},\
  \bibinfo {pages} {711} (\bibinfo {year} {2004})}\BibitemShut {NoStop}%
\bibitem [{\citenamefont {Richter}(1999)}]{Klausbook}%
  \BibitemOpen
  \bibfield  {author} {\bibinfo {author} {\bibfnamefont {K.}~\bibnamefont
  {Richter}},\ }\href@noop {} {\emph {\bibinfo {title} {{Semiclassical Theory
  of Mesoscopic Quantum Systems}}}}\ (\bibinfo  {publisher} {Springer},\
  \bibinfo {year} {1999})\BibitemShut {NoStop}%
\bibitem [{\citenamefont {Haake}(2010)}]{Libro_chaos}%
  \BibitemOpen
  \bibfield  {author} {\bibinfo {author} {\bibfnamefont {F.}~\bibnamefont
  {Haake}},\ }\href@noop {} {\emph {\bibinfo {title} {{Quantum Signatures of
  Chaos}}}}\ (\bibinfo  {publisher} {Springer-Verlag Berlin Heidelberg},\
  \bibinfo {year} {2010})\BibitemShut {NoStop}%
\bibitem [{\citenamefont {Whitney}\ \emph {et~al.}(2008)\citenamefont
  {Whitney}, \citenamefont {Jacquod},\ and\ \citenamefont
  {Petitjean}}]{Jacquod-transport}%
  \BibitemOpen
  \bibfield  {author} {\bibinfo {author} {\bibfnamefont {R.~S.}\ \bibnamefont
  {Whitney}}, \bibinfo {author} {\bibfnamefont {P.}~\bibnamefont {Jacquod}}, \
  and\ \bibinfo {author} {\bibfnamefont {C.}~\bibnamefont {Petitjean}},\ }\href
  {\doibase 10.1103/PhysRevB.77.045315} {\bibfield  {journal} {\bibinfo
  {journal} {Physical Review B}\ }\textbf {\bibinfo {volume} {77}},\ \bibinfo
  {pages} {045315} (\bibinfo {year} {2008})}\BibitemShut {NoStop}%
\bibitem [{\citenamefont {Waltner}(2012)}]{Daniel}%
  \BibitemOpen
  \bibfield  {author} {\bibinfo {author} {\bibfnamefont {D.}~\bibnamefont
  {Waltner}},\ }\href@noop {} {\emph {\bibinfo {title} {{Semiclassical Approach
  to Mesoscopic Systems}}}}\ (\bibinfo  {publisher} {Springer},\ \bibinfo
  {year} {2012})\BibitemShut {NoStop}%
\bibitem [{\citenamefont {Grabert}\ \emph {et~al.}(1988)\citenamefont
  {Grabert}, \citenamefont {Schramm},\ and\ \citenamefont {Ingold}}]{GRABERT}%
  \BibitemOpen
  \bibfield  {author} {\bibinfo {author} {\bibfnamefont {H.}~\bibnamefont
  {Grabert}}, \bibinfo {author} {\bibfnamefont {P.}~\bibnamefont {Schramm}}, \
  and\ \bibinfo {author} {\bibfnamefont {G.-L.}\ \bibnamefont {Ingold}},\
  }\href {\doibase 10.1016/0370-1573(88)90023-3} {\bibfield  {journal}
  {\bibinfo  {journal} {Physics Reports}\ }\textbf {\bibinfo {volume} {168}},\
  \bibinfo {pages} {115} (\bibinfo {year} {1988})}\BibitemShut {NoStop}%
\bibitem [{\citenamefont {Weiss}(2008)}]{Weiss}%
  \BibitemOpen
  \bibfield  {author} {\bibinfo {author} {\bibfnamefont {U.}~\bibnamefont
  {Weiss}},\ }\href {\doibase 10.1142/6738} {\emph {\bibinfo {title} {{Quantum
  Dissipative Systems}}}},\ \bibinfo {edition} {3rd}\ ed.\ (\bibinfo
  {publisher} {WORLD SCIENTIFIC},\ \bibinfo {year} {2008})\ \Eprint
  {http://arxiv.org/abs/https://www.worldscientific.com/doi/pdf/10.1142/6738}
  {https://www.worldscientific.com/doi/pdf/10.1142/6738} \BibitemShut {NoStop}%
\bibitem [{\citenamefont {{W. H. Zurek}}(1986)}]{Zurekinbook}%
  \BibitemOpen
  \bibfield  {author} {\bibinfo {author} {\bibnamefont {{W. H. Zurek}}},\
  }\href@noop {} {\emph {\bibinfo {title} {{Frontiers of Nonequilibrium
  Statistical Mechanics}}}},\ \bibinfo {edition} {g. t. moore, m. o. scully}\
  ed.\ (\bibinfo  {publisher} {Plenum Press},\ \bibinfo {year}
  {1986})\BibitemShut {NoStop}%
\bibitem [{\citenamefont {Waltner}\ \emph {et~al.}(2008)\citenamefont
  {Waltner}, \citenamefont {Guti{\'e}rrez}, \citenamefont {Goussev},\ and\
  \citenamefont {Richter}}]{Gutierrezprl}%
  \BibitemOpen
  \bibfield  {author} {\bibinfo {author} {\bibfnamefont {D.}~\bibnamefont
  {Waltner}}, \bibinfo {author} {\bibfnamefont {M.}~\bibnamefont
  {Guti{\'e}rrez}}, \bibinfo {author} {\bibfnamefont {A.}~\bibnamefont
  {Goussev}}, \ and\ \bibinfo {author} {\bibfnamefont {K.}~\bibnamefont
  {Richter}},\ }\href {\doibase 10.1103/PhysRevLett.101.174101} {\bibfield
  {journal} {\bibinfo  {journal} {Physical Review Letters}\ }\textbf {\bibinfo
  {volume} {101}},\ \bibinfo {pages} {174101} (\bibinfo {year}
  {2008})}\BibitemShut {NoStop}%
\bibitem [{\citenamefont {Kuipers}\ \emph {et~al.}(2009)\citenamefont
  {Kuipers}, \citenamefont {Waltner}, \citenamefont {Guti{\'e}rrez},\ and\
  \citenamefont {Richter}}]{Kuipers_2009}%
  \BibitemOpen
  \bibfield  {author} {\bibinfo {author} {\bibfnamefont {J.}~\bibnamefont
  {Kuipers}}, \bibinfo {author} {\bibfnamefont {D.}~\bibnamefont {Waltner}},
  \bibinfo {author} {\bibfnamefont {M.}~\bibnamefont {Guti{\'e}rrez}}, \ and\
  \bibinfo {author} {\bibfnamefont {K.}~\bibnamefont {Richter}},\ }\href
  {\doibase 10.1088/0951-7715/22/8/010} {\bibfield  {journal} {\bibinfo
  {journal} {Nonlinearity}\ }\textbf {\bibinfo {volume} {22}},\ \bibinfo
  {pages} {1945} (\bibinfo {year} {2009})}\BibitemShut {NoStop}%
\bibitem [{\citenamefont {Gutzwiller}(1990)}]{Gutzwiller}%
  \BibitemOpen
  \bibfield  {author} {\bibinfo {author} {\bibfnamefont {M.}~\bibnamefont
  {Gutzwiller}},\ }\href@noop {} {\emph {\bibinfo {title} {{Chaos in Classical
  and QuantumMechanics}}}}\ (\bibinfo  {publisher} {Springer},\ \bibinfo {year}
  {1990})\BibitemShut {NoStop}%
\bibitem [{\citenamefont {Sieber}(1999)}]{Sieber_1999}%
  \BibitemOpen
  \bibfield  {author} {\bibinfo {author} {\bibfnamefont {M.}~\bibnamefont
  {Sieber}},\ }\href {\doibase 10.1088/0305-4470/32/44/307} {\bibfield
  {journal} {\bibinfo  {journal} {Journal of Physics A: Mathematical and
  General}\ }\textbf {\bibinfo {volume} {32}},\ \bibinfo {pages} {7679}
  (\bibinfo {year} {1999})}\BibitemShut {NoStop}%
\bibitem [{\citenamefont {Case}(2008)}]{wigner}%
  \BibitemOpen
  \bibfield  {author} {\bibinfo {author} {\bibfnamefont {W.~B.}\ \bibnamefont
  {Case}},\ }\href {\doibase 10.1119/1.2957889} {\bibfield  {journal} {\bibinfo
   {journal} {American Journal of Physics}\ }\textbf {\bibinfo {volume} {76}},\
  \bibinfo {pages} {937} (\bibinfo {year} {2008})}\BibitemShut {NoStop}%
\bibitem [{\citenamefont {Polkovnikov}(2003)}]{PhysRevA-polk}%
  \BibitemOpen
  \bibfield  {author} {\bibinfo {author} {\bibfnamefont {A.}~\bibnamefont
  {Polkovnikov}},\ }\href {\doibase 10.1103/PhysRevA.68.053604} {\bibfield
  {journal} {\bibinfo  {journal} {Phys. Rev. A}\ }\textbf {\bibinfo {volume}
  {68}},\ \bibinfo {pages} {053604} (\bibinfo {year} {2003})}\BibitemShut
  {NoStop}%
\bibitem [{\citenamefont {Polkovnikov}(2010)}]{POLKOVNIKOV}%
  \BibitemOpen
  \bibfield  {author} {\bibinfo {author} {\bibfnamefont {A.}~\bibnamefont
  {Polkovnikov}},\ }\href {\doibase 10.1016/j.aop.2010.02.006} {\bibfield
  {journal} {\bibinfo  {journal} {Annals of Physics}\ }\textbf {\bibinfo
  {volume} {325}},\ \bibinfo {pages} {1790} (\bibinfo {year}
  {2010})}\BibitemShut {NoStop}%
\bibitem [{\citenamefont {Dujardin}\ \emph {et~al.}(2015)\citenamefont
  {Dujardin}, \citenamefont {Engl}, \citenamefont {Urbina},\ and\ \citenamefont
  {Schlagheck}}]{Truncated_W}%
  \BibitemOpen
  \bibfield  {author} {\bibinfo {author} {\bibfnamefont {J.}~\bibnamefont
  {Dujardin}}, \bibinfo {author} {\bibfnamefont {T.}~\bibnamefont {Engl}},
  \bibinfo {author} {\bibfnamefont {J.~D.}\ \bibnamefont {Urbina}}, \ and\
  \bibinfo {author} {\bibfnamefont {P.}~\bibnamefont {Schlagheck}},\ }\href
  {\doibase 10.1002/andp.201500113} {\bibfield  {journal} {\bibinfo  {journal}
  {Annalen der Physik}\ }\textbf {\bibinfo {volume} {527}},\ \bibinfo {pages}
  {629} (\bibinfo {year} {2015})}\BibitemShut {NoStop}%
\bibitem [{\citenamefont {Drummond}\ and\ \citenamefont
  {Opanchuk}(2017)}]{PhysRevA-Drummond}%
  \BibitemOpen
  \bibfield  {author} {\bibinfo {author} {\bibfnamefont {P.~D.}\ \bibnamefont
  {Drummond}}\ and\ \bibinfo {author} {\bibfnamefont {B.}~\bibnamefont
  {Opanchuk}},\ }\href {\doibase 10.1103/PhysRevA.96.043616} {\bibfield
  {journal} {\bibinfo  {journal} {Phys. Rev. A}\ }\textbf {\bibinfo {volume}
  {96}},\ \bibinfo {pages} {043616} (\bibinfo {year} {2017})}\BibitemShut
  {NoStop}%
\bibitem [{\citenamefont {Richter}\ and\ \citenamefont
  {Sieber}(2002)}]{Richter2002}%
  \BibitemOpen
  \bibfield  {author} {\bibinfo {author} {\bibfnamefont {K.}~\bibnamefont
  {Richter}}\ and\ \bibinfo {author} {\bibfnamefont {M.}~\bibnamefont
  {Sieber}},\ }\href {\doibase 10.1103/PhysRevLett.89.206801} {\bibfield
  {journal} {\bibinfo  {journal} {Physical Review Letters}\ }\textbf {\bibinfo
  {volume} {89}},\ \bibinfo {pages} {206801} (\bibinfo {year}
  {2002})}\BibitemShut {NoStop}%
\bibitem [{\citenamefont {Sieber}\ and\ \citenamefont
  {Richter}(2001)}]{Sieber2001}%
  \BibitemOpen
  \bibfield  {author} {\bibinfo {author} {\bibfnamefont {M.}~\bibnamefont
  {Sieber}}\ and\ \bibinfo {author} {\bibfnamefont {K.}~\bibnamefont
  {Richter}},\ }\href {\doibase 10.1238/Physica.Topical.090a00128} {\bibfield
  {journal} {\bibinfo  {journal} {Physica Scripta}\ }\textbf {\bibinfo {volume}
  {T90}},\ \bibinfo {pages} {128} (\bibinfo {year} {2001})}\BibitemShut
  {NoStop}%
\bibitem [{\citenamefont {Jacquod}\ and\ \citenamefont
  {Petitjean}(2009)}]{Pitetjean}%
  \BibitemOpen
  \bibfield  {author} {\bibinfo {author} {\bibfnamefont {P.}~\bibnamefont
  {Jacquod}}\ and\ \bibinfo {author} {\bibfnamefont {C.}~\bibnamefont
  {Petitjean}},\ }\href {\doibase 10.1080/00018730902831009} {\bibfield
  {journal} {\bibinfo  {journal} {Advances in Physics}\ }\textbf {\bibinfo
  {volume} {58}},\ \bibinfo {pages} {67} (\bibinfo {year} {2009})},\ \Eprint
  {http://arxiv.org/abs/https://doi.org/10.1080/00018730902831009}
  {https://doi.org/10.1080/00018730902831009} \BibitemShut {NoStop}%
\bibitem [{\citenamefont {Jacquod}\ and\ \citenamefont
  {Whitney}(2006)}]{Jacquod2006}%
  \BibitemOpen
  \bibfield  {author} {\bibinfo {author} {\bibfnamefont {P.}~\bibnamefont
  {Jacquod}}\ and\ \bibinfo {author} {\bibfnamefont {R.~S.}\ \bibnamefont
  {Whitney}},\ }\href {\doibase 10.1103/PhysRevB.73.195115} {\bibfield
  {journal} {\bibinfo  {journal} {Physical Review B}\ }\textbf {\bibinfo
  {volume} {73}},\ \bibinfo {pages} {195115} (\bibinfo {year}
  {2006})}\BibitemShut {NoStop}%
\bibitem [{\citenamefont {Guti{\'e}rrez}\ \emph {et~al.}(2009)\citenamefont
  {Guti{\'e}rrez}, \citenamefont {Waltner}, \citenamefont {Kuipers},\ and\
  \citenamefont {Richter}}]{Gutierrez2009}%
  \BibitemOpen
  \bibfield  {author} {\bibinfo {author} {\bibfnamefont {M.}~\bibnamefont
  {Guti{\'e}rrez}}, \bibinfo {author} {\bibfnamefont {D.}~\bibnamefont
  {Waltner}}, \bibinfo {author} {\bibfnamefont {J.}~\bibnamefont {Kuipers}}, \
  and\ \bibinfo {author} {\bibfnamefont {K.}~\bibnamefont {Richter}},\ }\href
  {\doibase 10.1103/PhysRevE.79.046212} {\bibfield  {journal} {\bibinfo
  {journal} {Physical Review E}\ }\textbf {\bibinfo {volume} {79}},\ \bibinfo
  {pages} {046212} (\bibinfo {year} {2009})}\BibitemShut {NoStop}%
\end{thebibliography}%
\end{document}